\documentstyle[preprint,aps]{revtex} 
\begin{document}
\preprint{ORNL-CTP-95-04 
and 
HEP-PH 9506270} 

\title{ Suppression of $\psi'$ and $J/\psi$ in High-Energy Heavy-Ion
Collisions}

\author{ Cheuk-Yin Wong }
\address{ Oak Ridge National Laboratory, Oak Ridge, TN 37831}

\date{\today}

\maketitle
 
\begin{abstract}

The experimental ratio of $\psi'$ to $J/\psi$ is approximately a
constant in $pA$ collisions, but decreases as the transverse energy
increases in nucleus-nucleus collisions.  These peculiar features can
be explained as arising from approximately the same $c \bar c$-baryon
absorption cross section for $\psi'$ and $J/\psi$ but greater
disruption probabilities for $\psi'$ than for $J/\psi$ due to the
interaction of the $c\bar c$ system with soft particles produced in
baryon-baryon collisions.

\end{abstract}


\narrowtext


High-energy heavy-ion collisions have become the focus of intense
research because of the possibility of producing a deconfined
quark-gluon plasma during such collisions \cite{QM93,Won94}. The
$J/\psi$ suppression has been suggested as a way to probe the
screening between a charm quark with its antiquark partner in the
plasma \cite{Mat86}.  While the $J/\psi$ suppression has been observed
\cite{Bag89,Abr92} as predicted, the phenomenon can be explained by the
absorption model \cite{Ger88,Vog91,Gav94}, which was actually
introduced earlier \cite{And77} to measure the total $\psi$-$N$ cross
section using $J/\psi$ suppression.  In the model of Gerschel and
H\"ufner \cite{Ger88}, the collision of a $J/\psi$ particle with
baryons of the colliding nuclei may lead to the breakup of the
$J/\psi$ into an open-charm pair.  An effective $J/\psi$-baryon
absorption cross section $\sigma_{\rm abs}(\psi N)$ of 5-6 mb can
explain the systematics of the suppression phenomenon
\cite{Ger88,Won94}. One can alternatively describe $J/\psi$
suppression in terms of the interaction of the $J/\psi$ particle with
produced hadrons (comovers) \cite{Vog91,Gav94}.  A comparison of the
production of $\psi'$ with $J/\psi$ has been suggested to
distinguish between deconfinement and absorption \cite{Gup92}.

The NA38 experimental measurements using protons and heavy ions at
200A GeV and 450A GeV reveal three features \cite{Lou95,Bag95}: 1)
$\psi'/\psi$ is approximately a constant in $pA$ collisions,
independent of energies, 2) $\psi'/\psi$ decreases as the transverse
energy $E_T^{0}$ increases in SU collisions, and 3) $\psi'/\psi$ for
SU collisions is about 0.5 of that for $pA$ collisions.  The first
feature, further supported by other $pA$ experiments \cite{Ald91},
implies that in $pA$ collisions $\psi'$ is suppressed in the same way
as $\psi$.  We would like to describe an absorption model with
$\sigma_{\rm abs}(\psi' N)=\sigma_{\rm abs}(\psi N)$ and with
additional soft-particle disruption to explain all three features of
the phenomenon.

The production of $J/\psi$ or $\psi'$ occurs by the interaction of the
partons of one baryon with the partons of the other baryon.  The
incipient $c\bar c$ pair is created with a radial dimension of the
order of $\sim 0.06$ fm at $t_{c \bar c}$.  It is necessary for the
incipient $c \bar c$ system to evolve to the bound state rms radius of
0.24 fm for $\psi$ at $t_{\psi}$ and 0.47 fm for $\psi'$ at
$t_{\psi'}$ \cite{Eic80,Bla87}.  Because $J/\psi$ is produced
predominantly in the central rapidity region \cite{Cra78}, the
incipient $c \bar c$ pair is produced predominantly in the central
rapidity region.

In soft particle production in a baryon-baryon collision, we envisage
Bjorken's inside-outside cascade picture \cite{Bjo73} or Webber's
picture of gluon branching \cite{Web84} as a $q$ and a $\bar q$ (or a
diquark) pull apart.  After the collision, the diquark of one nucleon
and the valence quark of the other nucleon pull apart and the gauge
field between them is polarized.  Gluons are emitted at $t_{ g}$, and
the system hadronizes at $t_h$.  The characteristics of these produced
gluons cannot yet be determined from nonperturbative QCD, but the
shape of the rapidity distribution of produced gluons should be close
to that of the produced hadrons.  Thus, produced gluons are found
predominantly in the central rapidity region.  Because we shall use
the $J/\psi$ (or $\psi'$) production rate in a nucleon-nucleon
collision as a unit of reference, it is not necessary to include
explicitly the interaction of the incipient $c\bar c$ system with
gluons and their hadronized products in the same nucleon-nucleon
collision, when we study $pA$ and nucleus-nucleus ($AB$) collisions.

The space-time diagram for a typical $pA$ collision is depicted
schematically in Fig.\ 1$a$.  The trajectory of an incipient $c \bar
c$ pair, which is produced predominantly in the central rapidity
region of the colliding baryons, does not cross the trajectories of
soft particles produced in earlier or later collisions.  Therefore,
there is little interaction between the produced $c \bar c$ system and
these soft particles.  However, the $c \bar c$ system collides with
baryons crossing its trajectory to lead to the breakup of the $c \bar
c$ system into $D \bar DX$.  Such a reaction requires the production
of at least one light-quark pair and is an inelastic process.  In the
collision at 200A GeV, the $c \bar c$ rapidities are separated from
the baryon rapidities by about two units and the reaction cross
section can be calculated in the Additive Quark Model (AQM)
\cite{Lev65}, whose approximate validity and connection with soft
Pomeron exchanges have been re-assessed recently from the systematics
of total hadron- and $\gamma$-hadron cross sections \cite{Don92}.
Using the Glauber theory and a Gaussian thickness function, the total
$c\bar c$-baryon inelastic cross section in the AQM is given by Eq.\
(12.27) of Ref.\ \cite{Won94}:
\begin{eqnarray} 
\label{eq:xsec}
\sigma_{\rm abs}({c\bar c-N })
=-2\pi \beta^2
\sum_{n=1}^{6} {6 \choose n} (- f)^n /n,
\end{eqnarray}
where $ f=\sigma_{cq}/\beta^2 ={\sigma_{cq} / 2\pi (\beta_{c \bar
c}^2+\beta_N^2 +\beta_{cq}^2) }\,,$ $\sigma_{cq}$ is the inelastic
cross section for the collision of $c$ (or $\bar c$) and a constituent
$q$ of the baryon, $\sqrt{3}\beta_{c \bar c}$ and $\sqrt{3} \beta_{N}$
are the rms radii of the $c \bar c$ and the baryon respectively, and
$\beta_{cq}$ is the $c$-$q$ interaction range.  We find below that
$\sigma_{\rm abs}(\psi N)=4.2$ mb.  Taking $\sqrt{3}\beta_N=0.74$ fm
\cite{Cho80} and neglecting $\beta_{cq}$, we obtain from Eq.\
(\ref{eq:xsec}) $\sigma_{cq}=0.753$ mb, which leads to $ \sigma_{\rm
abs}(\psi' N)=4.27$ mb, and $\sigma_{\rm abs}(c \bar c-N)=4.17$ mb for the
initial $c\bar c$ at $\sqrt{3}\beta_{c \bar c}=0.06$ fm. Thus, the
absorption cross section is approximate the same for any $c\bar c$
state during all stages of its evolution because $6 \sigma_{cq} <<
2\pi\beta^2 $.  Consequently, in $pA$ collisions, $\psi'$ is
suppressed in the same way as $\psi$ and $\psi'/\psi$ is a constant
independent of $A$ and collision energies, in agreement with
experimental observations.

The approximate equality of the absorption cross section for different
$c\bar c$ states is supported by the experimental ratio $\sigma_{\rm
total}(\psi' N)/\sigma_{\rm total}(\psi N)\sim 0.75$ to $0.86 \pm 0.15
$, for $\sqrt{s}$ ranging from 6.4 GeV to 21.7 GeV, as deduced from
the photoproduction of vector mesons \cite{Bri83,Won95b}.  The
approximate equality of the absorption cross section for the incipient
$c\bar c$ and other $c\bar c$ states implies that a small incipient $c
\bar c$ system is not transparent to the hadron medium, in agreement
with the absence of color transparency for small hadron systems as
indicated by experimental data in $A(e,e'p)$ reactions at high $Q^2$
\cite{Nei95}.

To study $AB$ collisions, we adopt a row-on-row picture and consider a
typical row with a cross section of the size of the nucleon-nucleon
inelastic cross section, $\sigma_{in}=29.4$ mb.  The space-time
diagram of the collision can be depicted schematically in Fig.\ 1$b$.
The trajectories of the $c \bar c$ system cross the trajectories of
colliding baryons, and the process of absorption due to the $c \bar
c$-baryon interaction is the same in $pA$ as in $AB$ collisions.
However, there are collisions, such as the ones at $E$ and $F$ in
Fig.\ 1$b$, where the trajectories of incipient $c\bar c$ systems
produced there cross the trajectories of the produced soft particles.
It is necessary to consider the additional interaction of the $c \bar
c$ systems with soft particles in $AB$ collisions but not in $pA$
collisions.  Because the $c \bar c$ systems and the gluons (or their
hadronized products) are produced predominantly in the central
rapidity region, their rapidities are not much separated and their
relative kinetic energies are not large.  At these low energies, the
absorption of a produced gluon by the $c\bar c$ system, the screening
of $c$ from $\bar c$ by gluons, and the inelastic gluon scattering
which excites the $c \bar c$ system to higher levels will contribute
to the breakup of the $c \bar c$ system.  As gluons carry color, the
cross sections for these breakup processes increase with the color
dipole moment of $c \bar c$ which is proportional to the separation
between $c$ and $\bar c$.  Furthermore, the threshold for $\psi'$
breakup is small compared to those for $\psi$ and $\chi_{1,2}$.  Thus,
the breakup probability for a $c\bar c(\psi')$ system due to $c\bar
c$-gluon interactions at low energies is greater than those for
$J/\psi$ and $\chi$.

The hadronized product (comovers) of the produced gluons can interact
with $J/\psi$ and $\psi'$ to lead to their breakup \cite{Vog91,Gav94}.
The breakup of $\psi'$, $\chi_{1,2}$ and $J/\psi$ into $D \bar D$
require the threshold energy of 52, $\sim 200$, and $640$ MeV
respectively.  In $c\bar c$-hadron interactions at low energies below
thresholds, $J/\psi$ and $\chi$ cannot be broken up by low energy
pions.  Above the thresholds, the interaction is mediated by color
gluon exchange which probes the color dipole moment of the $c\bar c$
system.  Thus, the breakup probability due to $c\bar c$-hadron
interactions at low energies is also larger for the $\psi'$ system
than those for the $J/\psi$ and the $\chi$ systems.  This is different from
the higher-energy (Pomeron-exchange dominated) Glauber case discussed
earlier where $\psi'$ and $J/\psi$ have about the same inelastic
baryon cross sections.

Assuming straight-line space-time trajectories and uniform
distribution of baryons in nuclei, we find 
\begin{eqnarray}
\label{eq:fin}
{ d \sigma_{{}_{\psi N}}^{{}^{AB}} (\bbox{b}) \over
\sigma_{{}_{\psi N}}^{{}^{NN}}~d\bbox{b} } = \int { d{\bbox{b}}_{{}_A}
\over \sigma_{\rm abs}^2(\psi N) } \biggl \{ 1 -\biggl [ 1-
T_{{}_A}(\bbox{b}_{{}_A}) \sigma_{\rm abs}(\psi N) \biggr ] ^A \biggr
\} \biggl \{ 1 -\biggl [ 1- T_{{}_B}(\bbox{b}-\bbox{b}_{{}_A})
\sigma_{\rm abs}(\psi N) \biggr ] ^B \biggr \} F(\bbox{b}_A)\,,
\end{eqnarray}
where $T_A(\bbox{b}_A)$ is the thickness function of $A$ 
and the disruption factor $F(\bbox{b}_A)$ is 
\begin{eqnarray}
\label{eq:fb}
F(\bbox{b}_A)={1 \over N_> N_<} \sum_{n=1}^{N_<} a(n) \sum_{i=1}^n
\exp\{-  \theta \sum_{j=1, j\ne i}^n ( k_{\psi g} t_{ij}^g  + k_{ \psi h} t_{ij}^h )  \} \,.
\end{eqnarray}
Here, $N_>(\bbox{b}_A)$ and $N_<(\bbox{b}_A)$ are the greater and the
smaller of the (rounded-off) numbers of target nucleons
$AT_A(\bbox{b}_A)\sigma_{in}$ and projectile nucleons
$BT_B(\bbox{b}-\bbox{b}_A)\sigma_{in}$ in the row at $\bbox{b}_A$ with
the cross section of $\sigma_{in}$, and $a(n)$ is
obtained by simple counting to be
\begin{eqnarray}
a(n)=2 {\rm~~for~~} n=1,2,...,N_<-1,~~{\rm
and}~~~a(N_<)=N_>-N_<+1\,.
\end{eqnarray}
In Eq.\ (\ref{eq:fb}), we have assumed that when a $c\bar c (\psi)$
system is produced in the $j$-th collision and soft particles are
produced at the same spatial location in the $i$-th collision, the
bound state survival probability is related to the $c\bar c$-gluon
interaction time $t_{ij}^g$ and the $c \bar c$-hadron interaction time
$t_{ij}^h$ by $e^{-\theta ( k_{\psi g} t_{ij}^g + k_{\psi
h}t_{ij}^h)}$ where $k_{\psi g}$ and $k_{\psi h}$ are rate constants,
averaged over the interaction history of the $c\bar c (\psi)$ system.
The interaction times, which must be nonnegative, are
\begin{eqnarray}
t_{ij}^g=t_i+t_h-{\rm Max}(t_i+t_g,t_j+t_{c\bar
c})\,,
\end{eqnarray} 
\begin{eqnarray}
\label{eq:fin6}
t_{ij}^h=t_n+t_f-{\rm Max}(t_i+t_h,t_j+t_{c \bar c} )\,,
\end{eqnarray} 
where $t_i=t_1+(i-1)2m_N d/\sqrt{s}$, $m_N$ is the nucleon mass,
$d(=1.93$ fm) is the inter-nucleon separation in a nucleus, $\sqrt{s}$
the nucleon-nucleon center-of-mass energy, and $t_f$ is the freeze-out
time. In Eq.\ (\ref{eq:fb}), the step function $ \theta=\Theta(A
T_{{}_A}(\bbox{b}_{{}_A}) \sigma_{in}-1) \Theta (B
T_{{}_B}(\bbox{b}-\bbox{b}_{{}_A})\sigma_{in} -1) \Theta(A-1)
\Theta(B-1) $ is introduced to insure that there is no soft-particle
disruption in $pA$ collisions.  The expressions for the production of
$\psi'$ can be obtained from Eqs.\ (\ref{eq:fin}-\ref{eq:fin6}) above
by changing $\psi$ into $\psi'$.  For simplicity, we do not treat
$\chi_{1,2}$ separately so that the extracted parameters are actually
for a ``$J/\psi$'' system with the observed $\psi:\chi=62:30$ mixture
\cite{Ant93}.

It is not yet possible to ascertain the exact nature of all
suppression mechanisms in $AB$ collisions because of the uncertainties
in the reaction cross sections (see below) and the characteristics of
produced gluons.  Besides the $c \bar c$-baryon absorption, the
suppression can be attributed to (A) produced gluons, (B) both
produced gluons and hadrons, (C) produced hadrons (as in the comover
model \cite{Vog91,Gav94}), or deconfined matter with no baryon
absorption \cite{Kha94}.  In our model, with $\sigma_{\rm abs}(\psi
N)=\sigma_{\rm abs}(\psi' N)=4.2$ mb fixed by $pA$ data and a set of
plausible time parameters $t_g=0.1 $, $t_h=1.2$, $t_f=3$, and
$t_{c\bar c}=0.06$ (in units of fm/c), we obtain results calculated
with rate constants (in c/fm) (A) $k_{\psi g}=0.2$, $k_{\psi' g}=3$,
$k_{\psi h}=k_{\psi' h}=0$ (gluon disruption only), (B) $k_{\psi
g}=0.2$, $k_{\psi' g}=1$, $k_{\psi h}=0$, $k_{\psi' h}=1$ (gluon
disruption for $J/\psi$ but gluon and hadron disruption for $\psi'$),
and (C) $k_{\psi g}=k_{\psi' g}=0$, $k_{\psi h}=0.12$, $k_{\psi' h}=3$
(hadron disruption only).  For the case when all impact parameters are
summed over, the quantity ${\cal B}\sigma_{{}_{J/\psi}}^{{}^{AB}}/AB$
is plotted as a function of $A^{1/3}+B^{1/3}$ in Fig.\ 2 for the three
cases considered.  The presence of additional soft-particle disruption
in $AB$ collisions relative to $pA$ collisions is consistent with the
experimental data in Fig.\ 2.  To study $\psi'/\psi$ data for SU
collisions at 200A GeV, we relate the transverse energy $E_T^0$
approximately to the impact parameter as given in Refs.\ {\cite{Bag90}
and \cite{Lou95}, and the ratio $\psi'/\psi$ is corrected for the
feeding of $J/\psi$ by $\psi'$.  The theoretical results for ${\cal
B}'\sigma(\psi')/{\cal B}\sigma(\psi)$ as a function of $E_T^0$ for
the cases of (A), (B), and (C) differ by about one percent and are
represented for simplicity by a single solid curve in Fig.\ 3.  The
theoretical ratio decreases with increasing $E_T^0$ and coincides with
the $pA$ limit, in good agreement with data.  When we include all
impact parameters, we obtain theoretically $[{\cal
B}'\sigma(\psi')/{\cal B}\sigma (J/\psi)]^{SU}/ [{\cal B}'\sigma
(\psi')/{\cal B}\sigma (J/\psi)]^{ pA} =0.62$ for the three sets of
parameters, approximately consistent with the experimental ratio
$[{\cal B}'\sigma(\psi')/{\cal B}\sigma (J/\psi)] ^{ SU }/ [{\cal
B}'\sigma(\psi')/{\cal B}\sigma (J/\psi)] ^{pA} =0.52 \pm 0.07$
~\cite{Lou95}.

The parameter sets in (A)-(C) suggest greater disruption for $\psi'$
than $J/\psi$ due to their interaction with soft-particles.  To
resolve the ambiguities, it is interesting to note that while
heavy-quark production by hadron-hadron collisions is inhibited by the
OZI rule, there is no such inhibition in gluon-gluon collisions.  The
fusion of energetic gluons produced in different baryon-baryon
collisions can lead to additional charm and strangeness production
\cite{Won95} and may explain the enhanced charm and dilepton
production in $AB$ collisions relative to $pA$ collisions observed in
\cite{Lou94}.

For a medium $m$ whose constituents move with an average velocity
$v_m$ with respect to $c\bar c(\psi_i)$, the rate constants $k_{\psi_i
m}$ is $\rho_m v_m \sigma_{\psi_i m} $ where $\rho_m=(dN_m/dy)/\pi
R_0^2 t_0$ is the contribution to the medium number density from a
single $NN$ collision and $\sigma_{\psi_i m}$ is the $\psi_i$-$m$
breakup cross section.  The time $t_0$ is the mean point in the time
when the medium exists in the specified form; $t_0=(t_g+t_h)/2=0.65$
fm/c for gluons, and $t_0=(t_h+t_f)/2=2.1$ fm/c for hadrons.  We can
estimate $\rho_m$ by taking $dN_h/dy=2 dN_g/dy=2.3$ and $R_0=0.5$ fm.
We can estimate $v_m$ by assuming that a produced gluon has a mass $M
\sim 1$ GeV \cite{Won95}.  At a temperature $T=200$ MeV which is often
found in these nuclear collisions, the most probable velocity is
$v_m=\sqrt{2kT/M}=0.6$ for gluons and $v_m \sim 1$ for hadrons.  Then,
the rate constants $k_{\psi_i m}$ suggest approximate
order-of-magnitudes of $\sigma_{\psi g} \sim 1.4$ mb, $\sigma_{\psi'
g} \sim 20$ mb for case (A), $\sigma_{\psi g} \sim 1.4$ mb,
$\sigma_{\psi' g} \sim \sigma_{\psi' h} \sim$ 7 mb for case (B), and
$\sigma_{\psi h} \sim 0.9$ mb, $\sigma_{\psi' h} \sim 21$ mb for case
(C).  The excessively large $\psi'$ cross sections required to explain
the $\psi'$ suppression in cases (A) and (C) may make the scenario (B)
tentatively a more attractive description.

The geometrical model \cite{Pov87} predicts the total hadron-$N$ cross
section at high energies proportional to the rms hadron radius and
suggests $\sigma_{\rm abs}(\psi'N)$ much larger than $\sigma_{\rm
abs}(\psi N)$, which is also assumed in the comover model
\cite{Gav94}.  Using PQCD, Kharzeev and Satz \cite{Kha94} claim that
$\psi N$ total cross section for $\sqrt{s}=6$ GeV is about 0.3 mb.
The geometrical model results and PQCD results differ from those
obtained here.  A recent calculation using an exchange potential gives
a $J/\psi$-hadron dissociation cross section about 7 mb at 0.8 GeV
kinetic energy \cite{Mar94}, in variance with the PQCD results of
Ref.\ \cite{Kha94}. Much work remains to be done to resolve the
differences.

\acknowledgements

The author would like to thank T.  C. Awes, F. E. Barnes, and Chun Wa
Wong for helpful discussions and Dr. C. Louren\c co for valuable
comments.  This research was supported by the Division of Nuclear
Physics, U.S. D. O. E.  under Contract DE-AC05-84OR21400 managed by
Lockheed Martin Energy Systems.

\begin{figure}[htbp]
\caption{ Schematic space-time diagram in the nucleon-nucleon
center-of-mass system, with the time axis pointing upward. ($a$) is for
a $pA$ collision and ($b$) is for an $AB$ collision.  The
trajectories of the baryons are given as solid lines, the trajectories
of an incipient $c\bar c$ system produced in some of the collisions
are represented by thick dashed lines and the trajectories of soft
particles produced in some of the baryon-baryon collisions by thin
dashed lines.  }
\label{fig1}
\end{figure}

\begin{figure}[htbp]
\caption{The quantity ${\protect\cal B} \sigma_{J/\psi}^{AB}/AB$ as a
function of $A^{1/3}+B^{1/3}$ for $pA$ and $AB$ collisions.  The data
points are from the NA3 Collaboration \protect\cite{Bad83} and the
NA38 Collaboration \protect\cite{Abr92}.  The solid curve gives
theoretical results for $pA$ collisions.  For $AB$ collisions, the
theoretical results are shown as the long-dashed curve for cases (A)
and (B) and as the dotted curve for case (C).  }
\label{fig2}
\end{figure}

\begin{figure}[htbp]
\caption{ The ratio ${\protect\cal B}'\sigma(\psi')/ {\protect\cal
B}\sigma(\psi)$ as a function of the transverse energy in SU
collisions at 200A GeV.  Data points are from
Ref. \protect\cite{Lou95}.  The theoretical results are shown as the
solid curve.  }
\label{fig3} 
\end{figure} 
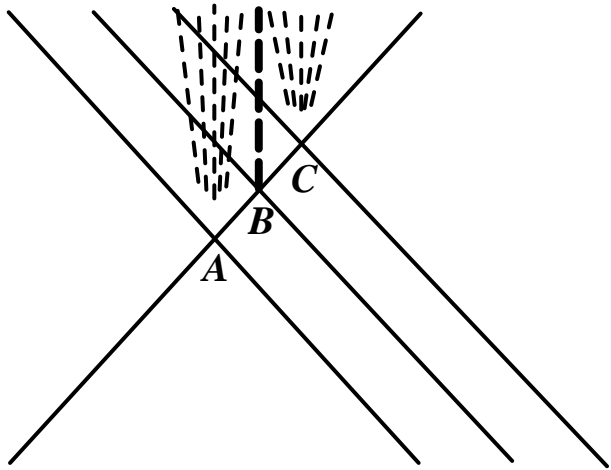
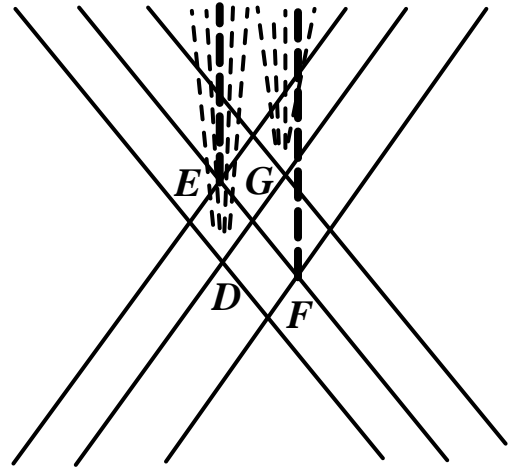
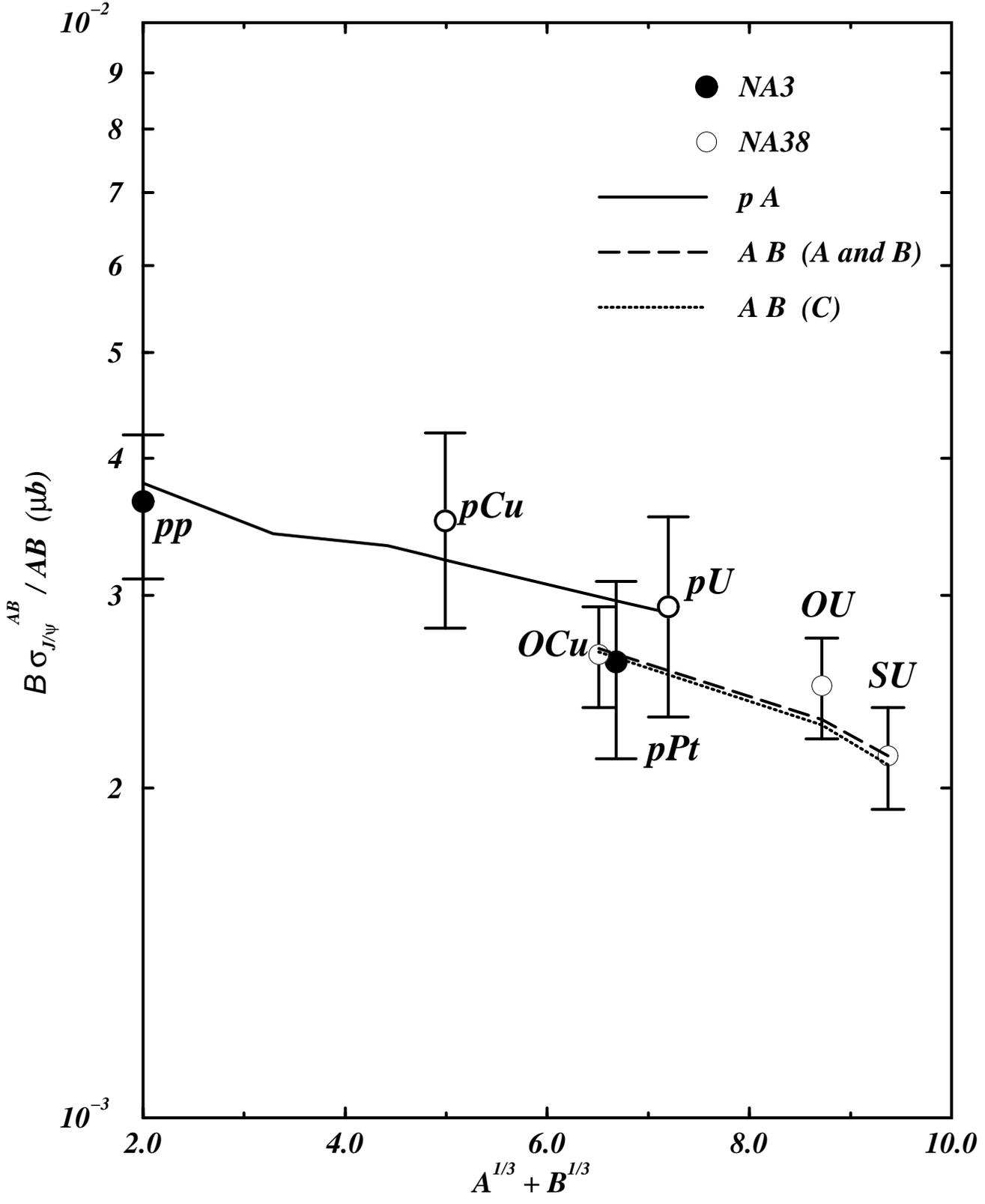
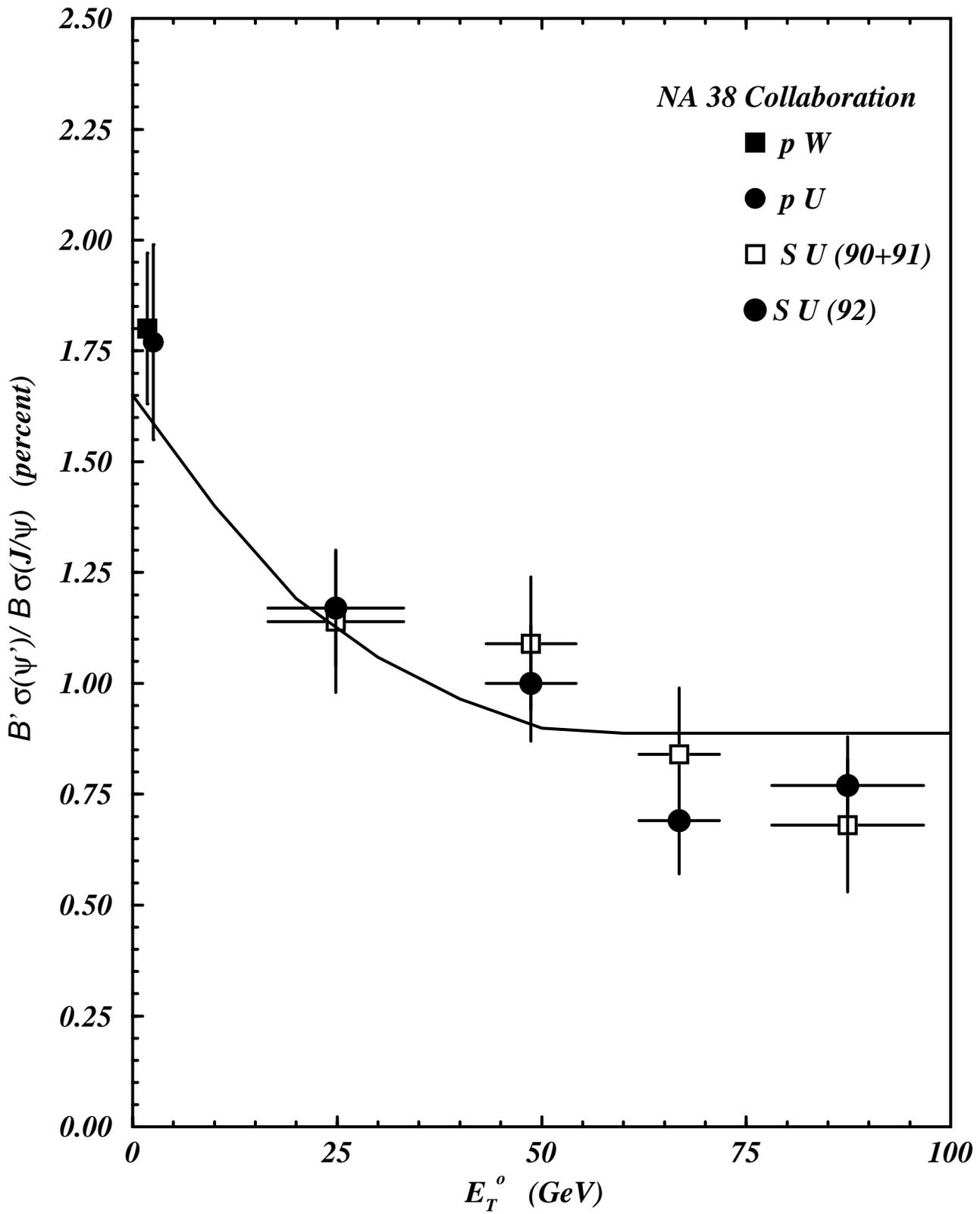
\end{document}